\documentclass{amsart}

\makeatletter

\providecommand{\LyX}{L\kern-.1667em\lower.25em\hbox{Y}\kern-.125emX\@}

 \theoremstyle{plain}    
 \newtheorem{thm}{Theorem}[section]
 \numberwithin{equation}{section} 
 \numberwithin{figure}{section} 
 \theoremstyle{plain}    
 \newtheorem{prop}[thm]{Proposition} 
 \theoremstyle{definition}
 \newtheorem{defn}[thm]{Definition}
 \theoremstyle{definition}
  \newtheorem{example}[thm]{Example}
 \theoremstyle{remark}
 \newtheorem{rem}[thm]{Remark}
 \theoremstyle{remark}    
 \newtheorem*{note*}{Note} 
 \theoremstyle{remark}    
 \newtheorem{notation}[thm]{Notation} 
 \theoremstyle{remark}    
 \newtheorem*{acknowledgement*}{Acknowledgement} 

\makeatother
\begin{document}

\title[Courant Algebroid-induced field theories]{AKSZ-BV Formalism and 
Courant algebroid-induced topological field
theories.}

\author{Dmitry Roytenberg}

\begin{abstract}
We give a detailed exposition of the Alexandrov-Kontsevich-Schwarz-Zaboronsky
superfield formalism using the language of graded manifolds. As a
main illustarting example, to every Courant algebroid structure we
associate canonically a three-dimensional topological sigma-model.
Using the AKSZ formalism, we construct the Batalin-Vilkovisky master
action for the model.
\end{abstract}
\maketitle

\section{Intro and Brief History.}

The standard procedure for quantizing classical field theories in
the Lagrangian approach is by using the Feynman path integral. From
the mathematical standpoint this is somewhat problematic, as it involves
``integration'' over the infinite-dimensional space of field configurations,
on which no sensible measure has been found to exist. Nevertheless,
the procedure can be made rigorous in the perturbative approach, provided
the classical theory does not have too many symmetries (``too many''
means, roughly speaking, an infinite-dimensional space). In the presence
of these gauge symmetries, however, the procedure needs to be modified,
as one has to integrate over the space of gauge-equivalence classes
of field configurations. This can be accomplished by gauge-fixing
(choosing a transversal slice to the gauge orbits), and in the late
60s Faddeev and Popov came up with an ingenious method of gauge-fixing
the path integral by introducing extra {}``ghost'' fields into the
action functional. These ghosts, corresponding to generators of the
gauge symmetries but of the opposite Grassman parity, were later incorporated
into the cohomological approach of Bechi-Rouet-Stora and Tyutin (BRST),
which is now considered the standard approach for quantizing gauge
theories. 

Unfortunately, the BRST approach fails in more complicated cases involving
so-called ``open'' algebras of symmetries (that is, when the symmetries
close under commutator bracket only modulo solutions of the classical
field equations). In the early 80s Batalin and Vilkovisky \cite{BV1}
developed a generalization of the BRST procedure which allows, in
principle, to handle symmetries of arbitrary complexity. The idea
is again to extend the field space by auxilliary fields (known as
higher-generation ghosts, antighosts and Lagrange multipliers), as
well as their ``conjugate antifields''. The field-antifield space
has two canonical structures. The first is an odd symplectic form,
such that the fields and their respective antifields are conjugate
with respect to the corresponding odd Poisson bracket \( (\cdot ,\cdot ) \)
(the ``antibracket''). The other is an odd second order differential
operator \( \Delta  \) (the so-called {}``BV Laplacian'') compatible
with the antibracket. The original action functional is extended to
a functional \( {\mathbf{S}} \) (called the master action) on this
odd symplectic supermanifold, obeying the so-called quantum master
equation \[
({\mathbf{S}},{\mathbf{S}})-2i\hbar \Delta {\mathbf{S}}=0\]
 The gauge-fixing is accomplished by choosing a Lagrangian submanifold,
and the perturbative expansion is computed by evaluating the path
integral over this Lagrangian submanifold. The generalized quantum
BRST operator is \( Q=-i\hbar \Delta +({\mathbf{S}},\cdot ) \). 

In practice the master action is computed by homological perturbation
theory which involves calculating relations among the generators of
the Euler-Lagrange ideal as well as the generators of the symmetries,
relations among the relations, etc. This is known in homological algebra
as the Koszul-Tate resolution, and can be very difficult to carry
out in general. But in mid-90s Alexandrov, Kontsevich, Schwarz and
Zaboronsky \cite{AKSZ} (referred to as AKSZ from now on) found a
simple and elegant procedure for constructing solutions to the \emph{classical
master equation}%
\footnote{which, for the cases they consider, also implies the quantum master
equation.
}: \[
({\mathbf{S}},{\mathbf{S}})=0\]
 Their approach uses mapping spaces of supermanifolds equipped with
additional structure. Here we use a slightly refined notion of \emph{differential
graded (dg) manifold,} which is a supermanifold equipped with a compatible
integer grading and a differential. The grading is needed to keep
track of the ghost number symmetry important in some applications.
If the source \( N \) is a dg manifold with an invariant measure,
and the target \( M \) is a dg symplectic manifold, then the space
of superfields \( \textrm{Maps}(N,M) \) acquires a canonical odd
symplectic structure; furthermore, any self-commuting hamiltonian
on \( M \) gives rise to a solution of the classical master equation.
In case \( N=T[1]N_{0} \) (corresponding to the algebra of differential
forms on a smooth manifold \( N_{0} \)) , one gets the master action
for topological field theories on \( N_{0} \) associated to various
structures on the target. In particular, AKSZ show that Witten's A
and B topological sigma-models are special cases of the AKSZ construction.

Cattaneo-Felder \cite{CF2} and Park \cite{Park} further refined
the AKSZ procedure by generalizing it to the case of manifolds with
boundary, and produced new examples: Cattaneo and Felder studied the
\emph{Poisson sigma-model} \cite{SchStr} on the disk \cite{CF1}
\cite{CF2}, while Park considered its higher-dimensional generalizations,
the \emph{topological open p-branes}. 

The Poisson sigma model is the most general 2d TFT that can be obtained
within the AKSZ framework (at least if the source is \( T[1]N_{0} \)),
the reason being that, for two-dimensional \( N_{0} \), the symplectic
form on the target must have degree 1, if ghost number symmetry is
to be preserved. This immediately implies that the target is of the
form \( M=T^{*}[1]M_{0} \), corresponding to the algebra of multivector
fields on a manifold \( M_{0} \), with the symplectic form corresponding
to the Schouten bracket. The diffrential structure on \( M \) is
then necessarily given by a self-commuting bivector field on \( M_{0} \),
i.e. a Poisson structure. The AKSZ procedure gives (the master action
for) the Poisson sigma model. Witten's A and B models can be obtained
as special cases of this, when the Poisson tensor is invertible and
there is a compatible complex structure.

Now, if we go one step further and consider three-dimensional \( N_{0} \),
the AKSZ formalism requires a symplectic form of degree 2 on the target,
and a self-commuting hamiltonian of degree 3. We have shown \cite{Roy4-GrSymp}
that such structures are in canonical 1-1 correspondence with what
is known as \emph{Courant algebroids}. A Courant algebroid is given
by specifying a bilinear operation on sections of a vector bundle
\( E\rightarrow M_{0} \) with an inner product, satisfying certain
natural properties. Thus the AKSZ procedure yields a canonical 3d
TFT associated to any Courant algebroid. Its classical action is given
by \[
S_{0}[X,A,F]=\int _{N_{0}}F_{i}dX^{i}+\frac{1}{2}A^{a}g_{ab}dA^{b}-A^{a}P_{a}^{i}(X)F_{i}+\frac{1}{6}T_{abc}(X)A^{a}A^{b}A^{c}\]
where the fields are the membrane world volume \( X:N_{0}\rightarrow M_{0} \),
an \( X^{*}E \)-valued 1-form \( A \) and an \( X^{*}T^{*}M_{0} \)-valued
2-form \( F \); \( g \) is the matrix of the inner product on \( E \),
while \( P_{a}^{i} \) and \( T_{abc} \) are the structure functions
of the Courant algebroid. This action has rather complicated symmetries,
requiring the introduction of ghosts for ghosts; the master action
contains terms up to degree 3 in antifields, involving up to 3rd derivatives
of the structure functions of the Courant algebroid. Known special
cases include the Chern-Simons theory (which is an ordinary gauge
theory) and Park's topological membrane.

The paper is organized as follows. Section 2 is a brief introduction
to the general theory of differential graded manifolds. Section 3
explains the AKSZ formalism in detail. Section 4 contains a discussion
of Courant algebroids and the associated closed membrane sigma model.

\begin{acknowledgement*}
This paper is based on the lectures given by the author at Rencontres
Mathematiques de Glanon in July 2003 to an audience of mathematicians
and mathematical physicists. I would like to thank the organizers
of the meeting and the people of Glanon for the excellent accomodation,
friendly atmosphere and hospitality. I would also like to thank A.
Cattaneo, A. Losev, J. Stasheff and T. Strobl for useful discussions
and comments.
\end{acknowledgement*}
\begin{note*}
After these notes had been completed, we learned that Ikeda \cite{Ikeda}
had obtained the Courant algebroid-valued sigma model by examining
consistent BV-deformations of the abelian Chern-Simons gauge theory
coupled with a zero-dimensional BF theory, rather than by immediate
application of AKSZ. Also, Hofman and Park \cite{HP1}\cite{HP2}
considered a generalization of the topological open membrane which
takes values in a quasi-Lie bialgebroid \cite{Roy3-QuasiLie}, which
is a Courant algebroid with a choice of a splitting. 
\end{note*}

\section{Differential graded manifolds.}

Here we collect the basic notions and fix the notation. The details
can be found, for instance, in \cite{Vor3} or \cite{Roy4-GrSymp}.

\subsection{Graded manifolds.}

\begin{defn}
A \emph{graded manifold} \( M \) over base \( M_{0} \) is a sheaf
of \( {\mathbb {Z}} \)-graded commutative algebras \( {\mathcal{C}}^{\cdot }(M) \)
over a smooth manifold \( M_{0} \) locally isomorphic to an algebra
of the form \( C^{\infty }(U)\otimes S^{\cdot }(V) \) where \( U\subset M \)
is an open set, \( V \) is a graded vector space, and \( S^{\cdot }(V) \)
is the free graded-commutative algebra on \( V \). Such a local isomorphism
is referred to as an \emph{affine coordinate chart} on \( M \); the
sheaf \( {\mathcal{C}}^{\cdot }(M) \) is called the \emph{sheaf of
polynomial functions} on \( M \).
\end{defn}
The generators of the algebra \( {\mathcal{C}}^{\cdot }(U)\simeq C^{\infty }(U)\otimes S^{\cdot }(V) \)
are viewed as local coordinates on \( M \). Coordinate transformations
are isomorphisms of algebras, hence in general non-linear. In what
follows we will be mostly concerned with \emph{nonnegatively} graded
manifolds. In this case the transformation law for a coordinate of
degree \( n \) is of the form \( x_{n}^{i}=A_{i'}^{i}(x_{0})x_{n}^{i'}+ \)
(terms in coordinates of lower degrees, of total degree \( n \)).
This explains the word {}``affine'' in the definition. 

We have decomposition \( {\mathcal{C}}^{\cdot }(M)=\oplus _{k}{\mathcal{C}}^{k}(M) \)
according to the degrees. Each \( {\mathcal{C}}^{k}(M) \) is a sheaf
of locally free \( C^{\infty }(M_{0}) \)-modules; in the nonnegative
case \( C^{\infty }(M_{0})={\mathcal{C}}^{0}(M) \). We denote by
\( {\mathcal{C}}_{k}(M) \) the subsheaf of algebras generated by
\( \oplus _{i\leq k}{\mathcal{C}}^{i}(M) \). These form a filtration
\[
\cdots \subset {\mathcal{C}}_{0}(M)\subset {\mathcal{C}}_{1}(M)\subset {\mathcal{C}}_{2}(M)\subset \cdots \]
 which in the nonnegative case is bounded below by \( {\mathcal{C}}_{0}(M)={\mathcal{C}}^{0}(M)=C^{\infty }(M_{0}) \).
In this case there is a corresponding tower of fibrations of graded
manifolds \begin{equation}
\label{tower}
M_{0}\leftarrow M_{1}\leftarrow M_{2}\leftarrow \cdots 
\end{equation}
 with \( M \) being the projective limit of the \( M_{n} \)'s. 

\begin{defn}
If \( M=M_{n} \) for some \( n \), we say \( \textrm{deg}(M)=n \).
Otherwise we say \( \textrm{deg}(M)=\infty  \).
\end{defn}
\begin{notation}
\label{shift} Given a vector bundle \( A\rightarrow M_{0} \), denote
by \( A[n] \) the graded manifold obtained by assigning degree \( n \)
to every fiber variable. The standard choice is \( n=1 \). Thus,
\( {\mathcal{C}}^{\cdot }(A[1]) \) is the sheaf of sections of \( \wedge ^{\cdot }A^{*} \)
with the standard grading. 
\end{notation}
\begin{example}
For \( A=TM_{0} \) we have graded manifolds \( T[1]M_{0} \) corresponding
to the sheaf of differential forms on \( M_{0} \), and \( T^{*}[1]M_{0} \)
corresponding to the sheaf of multi-vector fields. In general, the
graded manifold \( M_{1} \) in (\ref{tower}) is always of the form
\( A[1] \) for some \( A \), whereas the fibrations \( M_{k+1}\rightarrow M_{k} \)
for \( k>0 \) are in general affine rather than vector bundles. 
\end{example}
When working with graded manifolds it is very convenient to use the
\emph{Euler vector field} \( \epsilon  \), which is defined as the
derivation of \( {\mathcal{C}}^{\cdot }(M) \) such that \( \epsilon (f)=kf \)
if \( f\in {\mathcal{C}}^{k}(M) \). In a local affine chart \( \epsilon =\sum _{\alpha }\textrm{deg}(x^{\alpha })x^{\alpha }\frac{\partial }{\partial x^{\alpha }} \).
In particular, \( M_{0} \) is recovered as the set of fixed points
of \( \epsilon  \), which then acts on the normal bundle of \( M_{0} \)
in \( M \); \( \textrm{deg}(M) \) defined above is simply the highest
weight of this action, {}``the highest degree of a local coordinate''.

Graded manifolds form a category \( GrMflds \) with \( \textrm{Hom}(M,N)=\textrm{Hom}({\mathcal{C}}^{\cdot }(N),{\mathcal{C}}^{\cdot }(M)) \)
in graded algebras. Any smooth manifold is a graded manifold with
\( \epsilon =0 \); this gives a fully faithful embedding into \( GrMflds \).
Furthermore, the assignment \( A\mapsto A[1] \) gives a fully faithful
embedding of the category \( Vect \) of vector bundles into \( GrMflds \). 

One also has the forgetful functor into the category \( SuperMflds \)
of supermanifolds, which only remembers the grading modulo \( 2 \).
The algebra \( {\mathcal{C}}^{\cdot }(M) \) is completed by allowing
arbitrary smooth functions of all even variables. The Euler vector
field on \( M \) is related to the parity operator on the corresponding
supermanifold by \( P=(-1)^{\epsilon } \).

\begin{rem}
The roles of the \( {\mathbb {Z}}_{2} \)-grading (parity) and the
\( {\mathbb {Z}} \)-grading are actually quite different. The former
is responsible for the signs in formulas and in physics distinguishes
bosons from fermions. The latter (the {}``ghost number'' grading
in physics) distinguishes physical fields from auxilliary ones (ghosts,
antifields, etc.); for us the (nonnegative) integer grading has the
added advantage of imposing rigid structure. But in general these
two gradings are independent from one another \cite{Vor3}. Our requirement
that they be compatible, aside from simplifying the presentation somewhat,
is based on the fact that the classical field theories we consider
here are bosonic. But in principle the AKSZ-BV formalism can handle
more general bosonic-fermionic theories.
\end{rem}

\subsection{Vector bundles.}

The notion of a vector bundle in \( GrMflds \) can be defined in
two ways. Given a graded manifold \( M \), we can consider a sheaf
of locally free \( {\mathcal{C}}^{\cdot }(M) \)-modules. Alternatively,
we can consider a graded manifold \( A \) with a surjective submersion
\( A\rightarrow M \) in \( GrMflds \) and a linear structure on
\( A \) given by an additional Euler vector field \( \epsilon _{\textrm{vect}} \)
which assigns degree \( 1 \) to every fiber variable and degree \( 0 \)
to all functions pulled back from \( M \) (consequently, \( \epsilon _{\textrm{vect}} \)
necessarily commutes with the Euler vector field \( \epsilon _{A} \)
defining the grading on \( A \)). The \( {\mathcal{C}}^{\cdot }(M) \)-module
of sections of \( A \) can then be recovered by a variant of the
mapping space construction described in the next section. One gets
the following generalization of \ref{shift}:

\begin{notation}
Given a vector bundle \( A\rightarrow M \) in \( GrMflds \), denote
by \( A[n] \) the graded manifold corresponding to the Euler vector
field \( \epsilon _{A}+n\epsilon _{\textrm{vect}} \). It is again
a vector bundle over \( M \) (with the same \( \epsilon _{\textrm{vect}} \)). 
\end{notation}
\begin{example}
For a graded manifold \( M \), \( TM \) and \( T^{*}M \) are vector
bundles in \( GrMflds \), with the Euler vector field \( \epsilon _{M} \)
lifting canonically via Lie derivative (see below). Thus, we also
get vector bundles \( T[n]M \) and \( T^{*}[n]M \) for various integers
\( n \). For instance, in \( T^{*}[n]M \) one has \( \textrm{deg}(p_{\alpha })+\textrm{deg}(q^{\alpha })=n \).
Iterating this construction gives rise to many interesting graded
manifolds, such as \( T^{*}[2]T^{*}[1]M_{0} \) and so on.
\end{example}

\subsection{Mapping spaces.}

Because the structure sheaf \( {\mathcal{C}}^{\cdot }(M) \) of a
graded manifold can contain nilpotents, care must be taken in defining
such notions as points, maps, sections of vector bundles and so on.
The most general solution is to use the categorical approach which,
as far as we know, goes back to Grothendieck.

We shall assume that for any smooth manifolds \( M_{0} \), \( N_{0} \)
the space of smooth maps \( \textrm{Maps}(M_{0},N_{0}) \) is endowed
with a smooth structure making it an infinite-dimensional manifold.

\begin{prop}
Fix graded manifolds \( M \) and \( N \). Then the functor from
\( GrMflds \) to \( Sets \) given by \( Z\mapsto \textrm{Hom}(N\times Z,M) \)
is representable. In other words, there exists a graded manifold \( \textrm{Maps}(N,M) \),
unique up to a unique isomorphism, such that \( \textrm{Hom}(N\times Z,M)=\textrm{Hom}(Z,\textrm{Maps}(N,M)) \).
Its base \( \textrm{Maps}(N,M)_{0} \) is \( \textrm{Hom}(N,M) \),
viewed as an infinite-dimensional smooth manifold containing \( \textrm{Maps}(N_{0},M_{0}) \). 
\end{prop}
\begin{proof}
(Sketch) For simplicity assume that \( M \) and \( N \) (but not
\( Z \)!) are nonnegatively graded, and \( \textrm{deg}(N)=1 \)
(it will become clear how to handle the general case). Let \( x=\{x_{0},x_{1}\} \)
be coordinates on \( N \), \( y=\{y_{0},y_{1},y_{2},\ldots \} \)
(subscripts indicate the degree) on \( M \), \( z \) on \( Z \)
(the latter are viewed as parameters). Then any morphism from \( N\times Z \)
to \( M \) in \( GrMflds \) has a coordinate expression of the form\[
\begin{array}{ccccl}
y_{0} & = & y_{0}(x,z) & = & y_{0,0}(x_{0},z)+y_{0,-1}(x_{0},z)x_{1}+\frac{1}{2}y_{0,-2}(x_{0},z)x_{1}^{2}+\cdots \\
y_{1} & = & y_{1}(x,z) & = & y_{1,1}(x_{0},z)+y_{1,0}(x_{0},z)x_{1}+\frac{1}{2}y_{1,-1}(x_{0},z)x_{1}^{2}+\cdots \\
y_{2} & = & y_{2}(x,z) & = & y_{2,2}(x_{0},z)+y_{2,1}(x_{0},z)x_{1}+\frac{1}{2}y_{2,0}(x_{0},z)x_{1}^{2}+\cdots \\
\ldots  & \ldots  & \ldots  & \ldots  & \ldots 
\end{array}\]
 Here we suppress the running indices, so that for instance \( \frac{1}{2}y_{0,-2}x_{1}^{2} \)
actually means \( \frac{1}{2}y^{i}_{0,-2,\mu \nu }x_{1}^{\mu }x_{1}^{\nu } \).
The coefficients are arbitrary expressions in \( x_{0} \) and \( z \)
of total degree indicated by the second subscript. Now we let the
arbitrary functions of \( x_{0} \), \( y_{p,q}(x_{0}) \) (viewed
as coordinates of degree \( q \)), parametrize \( \textrm{Maps}(N,M) \).
As the parameter space \( Z \) is completely arbitrary, the transformation
rules for the \( y_{p,q} \)'s are determined by those for the \( x \)'s
and \( y \)'s. The universal property is immediately verified: it's
just a matter of tautologically rewriting \( y_{p,q}(x_{0},z) \)
as \( y_{p,q}(x_{0})(z) \). The degree \( 0 \) coordinates \( y_{p,0}(x_{0}) \)
parametrize the {}``actual'' (i.e. degree-preserving) maps from
\( N \) to \( M \).
\end{proof}
\begin{rem}
If \( N \) is just a point, we recover \( M \) as \( \textrm{Maps}(\textrm{pt},M) \).
This shows the correct way to think of points in graded manifolds.
In general, when dealing with points, maps, sections and so on, one
must allow them to implicitly depend on arbitrary additional parameters.
\end{rem}

\begin{rem}
In physics, the maps such as \( y_{p,0} \) appear as various kinds
of physical fields, whereas the non-zero degree \( y_{p,q} \)'s are
referred to as {}``ghosts'', {}``antifields'' and so on, and considered
as non-physical, auxilliary fields. Correspondingly, the degree \( q \)
in physics is called {}``ghost number''. Expressions such as \( y_{p}(x)=\sum y_{p,p-k}(x_{0})x_{1}^{k} \)
are called \emph{superfields}. In view of the above Proposition, there
is a good reason (apart from the physical considerations) to call
the non-zero degree \( y_{p,q} \)'s {}``ghosts'': they only appear
when additional parameters are introduced!
\end{rem}

\begin{rem}
We shall not have a detailed discussion of graded (Lie) groups, but
it should be clear how to define them, especially for the reader familiar
with supergroups, on which there is by now extensive literature. Graded
groups are just group objects in the category \( GrMflds \); the
group axioms are expressed by commutative diagrams. In particular,
for a graded manifold \( M \), \( \textrm{Diff}(M) \) is a graded
group, constructed just as in the above Proposition. The direct product
\( \textrm{Diff}(N)\times \textrm{Diff}(M) \) acts on \( \textrm{Maps}(N,M) \)
in an obvious way, just like for ordinary manifolds.
\end{rem}
\begin{example}
The following statements are easily verified:
\begin{enumerate}
\item \( \textrm{Maps}(S^{1},{\mathbb {R}}[1])\simeq {\mathbb {R}}^{\infty }[1] \)
(with Fourier modes as coordinates).
\item \( \textrm{Maps}({\mathbb {R}}[-1],M)\simeq T[1]M \) for any graded
manifold \( M \). 
\end{enumerate}
\end{example}

\subsection{Differentials.}

A vector field on a graded manifold \( M \) is by definition a derivation
of  \( {\mathcal{C}}^{\cdot }(M) \). Vector fields form a sheaf of
graded Lie algebras under the graded commutator.

\begin{defn}
A \emph{differential graded manifold} is a graded manifold \( M \)
with a self-commuting vector field \( Q \) of degree \( +1 \), i.e.
\( [\epsilon ,Q]=Q \) and \( [Q,Q]=2Q^{2}=0 \).
\end{defn}
\begin{example}
\( \textrm{Vect}({\mathbb {R}}[-1]) \) is spanned by the Euler vector
field \( \epsilon _{0}=-\theta \frac{d}{d\theta } \) and the differential
\( Q_{0}=\frac{d}{d\theta } \), with commutation relations as in
above definition. It acts on \( \textrm{Maps}({\mathbb {R}}[-1],M)\simeq T[1]M \)
in an obvoius way, giving rise to the grading of differential forms
and the de Rham differential \( d=dx^{i}\frac{\partial }{\partial x^{i}} \).
In general, any differential graded manifold is, by definition, acted
upon by the graded group \( \textrm{Diff}({\mathbb {R}}[-1]) \).
The differential integrates to an action of the subgroup \( {\mathbb {R}}[1] \)
(the {}``odd time flow'').
\end{example}

\begin{example}
Any vector field \( v=v^{a}(x)\frac{\partial }{\partial x^{a}} \)
on a graded manifold \( M \) gives rise to a vector field \( \iota _{v}=(-1)^{\textrm{deg}(v)}v^{a}(x)\frac{\partial }{\partial dx^{a}} \)
on \( T[1]M \), and consequently also the Lie derivative \( L_{v}=[\iota _{v},d] \).
One has \( \textrm{deg}(\iota _{v})=\textrm{deg}(v)-1 \), \( \textrm{deg}(L_{v})=\textrm{deg}(v) \)
and \( L_{[v,w]}=[L_{v},L_{w}] \).
\end{example}
The assignment \( M_{0}\mapsto (T[1]M_{0},d) \) is a full and faithful
functor from \( Mflds \) to \( dgMflds \). It is the \emph{right}
adjoint to the forgetful functor which assigns to any dg manifold
\( N \) its base \( N_{0} \). In particular, the unit of adjunction
gives a canonical dg map \( (M,Q)\rightarrow (T[1]M_{0},d) \) for
any dg manifold \( M \) with base \( M_{0} \), called the \emph{anchor}. 

There are many interesting dg manifolds around. Those that are (as
graded manifolds) of the form \( A[1] \) for some vector bundle \( A\rightarrow M_{0} \)
are otherwise known as \emph{Lie algebroids}. Those that come from
Courant algebroids (see next section), however, are not of this form.

\subsection{Symplectic and Poisson structures.}

Recall that the graded manifold structure on \( T[1]M \) is given
by the Euler vector field \( \epsilon _{\textrm{tot }}=L_{\epsilon }+\epsilon _{\textrm{vect}} \)
where \( \epsilon  \) gives the grading on \( M \) and \( L \)
denotes the Lie derivative. However, we shall speak of the degree%
\footnote{Physicists would prefer the term {}``ghost number''.
} of a differential form on \( M \) meaning only the action of the
induced Euler vector field \( L_{\epsilon } \), rather than \( \epsilon _{\textrm{tot }} \)
or \( \epsilon _{\textrm{vect}} \). Thus, {}``a \( p \)-form \( \omega  \)
of degree \( q \)'' means \( \epsilon _{\textrm{vect}}\omega =p\omega  \)
and \( L_{\epsilon }\omega =q\omega  \). In particular, a \emph{symplectic
structure of degree \( n \)} is a closed non-degenerate two-form
\( \omega  \) on \( M \) with \( L_{\epsilon }\omega =n\omega  \). 

Given such a symplectic structure, one defines for any function \( f\in {\mathcal{C}}^{\cdot }(M) \)
its hamiltonian vector field \( X_{f} \) (of degree \( \textrm{deg}(X_{f})=\textrm{deg}(f)-n \))
via \[
\iota _{X_{f}}\omega =(-1)^{n+1}df\]
 and for any two functions \( f,g \) their Poisson bracket \[
\{f,g\}=X_{f}\cdot g=(-1)^{\textrm{deg}(X_{f})}\iota _{X_{f}}dg=(-1)^{\textrm{deg}(f)+1}\iota _{X_{f}}\iota _{X_{g}}\omega \]
 It's easily verified that the bracket defines a graded Lie algebra
structure on \( {\mathcal{C}}^{\cdot }(M)[n] \) and \( f\mapsto X_{f} \)
is a homomorphism%
\footnote{Strictly speaking, one should write \( f\mapsto X_{f[-n]} \) 
}. 

\begin{example}
The basic example of a graded symplectic manifold is \( T^{*}[n]M \)
for some graded manifold \( M \) and an integer \( n \). The symplectic
form is canonical \( \omega =(-1)^{\tilde{\omega }\tilde{\alpha }}dp_{\alpha }dq^{\alpha } \)
where \( \tilde{\omega }=n\; \textrm{mod }2 \) and \( \tilde{\alpha }=\textrm{deg}(q^{\alpha })\; \textrm{mod }2 \).
The signs are chosen so that we have always \( \{p_{\alpha },q^{\beta }\}=\delta _{\alpha }^{\beta } \).
For \( n=1 \) and \( M=M_{0} \) an ordinary manifold, the Poisson
bracket induced by \( \omega  \) is just the Schouten bracket of
multivector fields; it is easy to show that any nonnegatively graded
symplectic manifold of degree \( 1 \) is canonically isomorphic to
\( T^{*}[1]M_{0} \) for some ordinary manifold \( M_{0} \).
\end{example}

\begin{example}
If \( V \) is a vector space, any nondegenerate symmetric bilinear
form on \( V \) can be viewed as a symplectic structure on \( V[1] \)
as follows: \( \omega =\frac{1}{2}d\xi ^{a}g_{ab}d\xi ^{b} \) where
\( g_{ab}=<e_{a},e_{b}> \)for some basis \( \{e_{a}\} \) of \( V \).
Clearly this \( \omega  \) has degree \( 2 \).
\end{example}
The following statements concerning a graded symplectic manifold \( (M,\omega ) \)
are easily verified:

\begin{enumerate}
\item If \( M \) is nonnegatively graded, \( \textrm{deg}(M)\leq \textrm{deg}(\omega ) \). 
\item If \( \textrm{deg}(\omega )=n\neq 0 \), then \( \omega =d\alpha  \),
where \( \alpha =\frac{1}{n}\iota _{\epsilon }\omega  \).
\item If \( v \) is a vector field of degree \( m\neq -n \), such that
\( L_{v}\omega =0 \), then \( \iota _{v}\omega =\pm d(\frac{1}{m+n}\iota _{v}\iota _{\epsilon }\omega ) \).
\end{enumerate}
A corollary of the last statement is that for a graded symplectic
manifold of degree \( n\neq -1 \), any differential preserving \( \omega  \)
is of the form \( Q=\{\Theta ,\cdot \} \) for some \( \Theta \in {\mathcal{C}}^{n+1}(M) \)
obeying the Maurer-Cartan equation \( \{\Theta ,\Theta \}=0 \). A
graded manifold equipped with such an \( \omega  \) and \( \Theta  \)
will be referred to as a \emph{differential graded symplectic manifold.}
For \( n=1 \) and \( M=T^{*}[1]M_{0} \) this is the same as an ordinary
Poisson structure on \( M_{0} \).

\subsection{Measure and integration.}

The general integration theory on supermanifolds is quite nontrivial.
Fortunately, all we need here is the notion of the integral of a function
over the whole graded manifold, i.e. a volume form or a measure. By
a measure on \( M \) we shall understand a functional on the space
of compactly supported functions (denoted by \( f\mapsto \int _{M}\mu f \))
such that locally \( \mu  \) is the Berezinian measure of the form
\( F(x)dx \). We call \( \mu  \) \emph{nondegenerate} if the bilinear
form \( <f,g>=\int _{M}\mu fg \) is. Given a vector field \( v \)
on \( M \) we say a measure \( \mu  \) on \( M \) is \emph{\( v \)-invariant}
if \( \int _{M}\mu (vf)=0 \) for any \( f \). 

Given two graded manifolds \( M \) and \( N \) and a measure \( \mu  \)
on \( N \) one can define the push-forward or fiber integration of
differential forms. This is a chain map \( \mu _{*}:\Omega ^{k}(N\times M)\rightarrow \Omega ^{k}(M)[\textrm{deg}\mu ] \)
defined as follows: \[
\mu _{*}\omega (y)(v_{1},\ldots ,v_{k})=\int _{N}\mu (x)\omega (x,y)(v_{1},\ldots ,v_{k})\]
 If \( \mu  \) is \( v \)-invariant for some vector field \( v \)
on \( N \), one has \( \mu _{*}L_{v_{1}}=0 \), where \( v_{1} \)
is the lift to \( N\times M \) of \( v \).

\begin{example}
If \( N_{0} \) is a closed oriented smooth \( n \)-manifold, the
graded manifold \( T[1]N_{0} \) has a canonical measure defined as
follows: \[
\int _{T[1]N_{0}}\mu f=\int _{N_{0}}f^{\textrm{top}}\]
 where \( f^{\textrm{top}} \) denotes the top-degree component of
the inhomogeneous differential form \( f \). This measure has degree
\( -n \), is invariant with respect to the de Rham vector field \( d \)
(Stokes' Theorem) and also with respect to all vector fields of the
form \( \iota _{v} \) (since the top degree component of \( \iota _{v}f \)
is always zero); hence, it is invariant with respect to all Lie derivatives
\( L_{v} \), i.e. all isotopies of \( N_{0} \) (in fact, all orientation-preserving
diffeomorphisms). The induced nondegenerate pairing of differential
forms is the Poincare pairing.
\end{example}

\section{The AKSZ formalism.}

Let us fix the following data.

\textbf{The source}: a dg manifold \( (N,D) \) endowed with a nondegenerate
\( D \)-invariant measure \( \mu  \) of degree \( -n-1 \) for a
positive integer \( n \). In practice we will consider \( N=T[1]N_{0} \)
for a closed oriented \( (n+1) \)-manifold \( N_{0} \), with \( D=d \),
the de Rham vector field, and \( \mu  \) the canonical measure.

\textbf{The target}: a dg symplectic manifold \( (M,\omega ,Q) \)
with \( \textrm{deg}(\omega )=n \). Then \( Q \) is of the form
\( \{\Theta ,\cdot \} \) for a solution \( \Theta \in {\mathcal{C}}^{n+1}(M) \)
of the Maurer-Cartan equation: \[
\{\Theta ,\Theta \}=0\]
 We shall assume here that both \( M \) and \( N \) are nonnegatively
graded. Our space of superfields will be \( {\mathcal{P}}=\textrm{Maps}(N,M) \).
This is a graded manifold, with the degree of a functional referred
to as the \emph{ghost number}. In the algebra of functions on \( {\mathcal{P}} \),
generators of non-negative ghost number will be called \emph{fields},
those of negative ghost number -- the \emph{antifields}. Among the
fields we further distinguish between the \emph{classical fields}
(of ghost number zero) and \emph{ghosts}, ghosts for ghosts, etc.
(of positive ghost number). We shall put a \( QP \)-structure on
\( {\mathcal{P}} \), i.e. an odd symplectic form \( \Omega  \) of
degree \( -1 \) and a homological vector field (the BRST differential)
\( {\mathbf{Q}} \) of degree \( +1 \) preserving \( \Omega  \).
It is not guaranteed in general that such a field is hamiltonian (since
\( {\mathbf{Q}} \) and \( \Omega  \) are of opposite degree), but
in our case it will be since \( \Omega  \) is exact. Thus we will
obtain a solution \( {\mathbf{S}} \) of the classical Batalin-Vilkovisky
master equation: \[
{(}{\mathbf{S}},{\mathbf{S}}{)}=0\]
 where \( {(}\cdot ,\cdot {)} \) is the odd Poisson bracket corresponding
to \( {\omega } \), otherwise known as the \emph{BV antibracket},
while \( {\mathbf{S}} \) (of ghost number zero) is the hamiltonian
of \( {\mathbf{Q}} \). 

Let us first define the \( Q \)-structure. As remarked above, the
graded group \( \textrm{Diff}(N)\times \textrm{Diff}(M) \) acts naturally
on \( {\mathcal{P}} \). Therefore, the vector fields \( D \) on
\( N \) and \( Q \) on \( M \) induce a pair of commuting homological
vector fields on \( {\mathcal{P}} \) which we denote by \( \hat{D} \)
and \( \check{Q} \), respectively. Hence, any linear combination
\( {\mathbf{Q}}=a\hat{D}+b\check{Q} \) is a \( Q \)-structure. We
shall fix the coefficients later to get the master action in the form
we want. 

Now for the \( P \)-structure. There is a canonical evaluation map
\( \textrm{ev}:N\times {\mathcal{P}}\rightarrow M \) given by \( (x,f)\mapsto f(x) \).
This enables us to pull back differential forms on \( M \) using
\( \textrm{ev} \), and then push them forward to \( {\mathcal{P}} \)
using the measure \( \mu  \). The resulting chain map \( \mu _{*}\textrm{ev}^{*}:\Omega ^{k}(M)\rightarrow \Omega ^{k}({\mathcal{P}}) \)
has ghost number \( \textrm{deg}(\mu )=-n-1 \). So let us define
\( \Omega =\mu _{*}\textrm{ev}^{*}\omega  \). If the measure \( \mu  \)
is nondegenerate, this defines a \( P \)-structure. 

It remains to check that the \( P \) and \( Q \)-structures are
compatible and calculate the master action. We first observe that
\( \mu _{*}\textrm{ev}^{*} \) applied to functions preserves the
Poisson brackets. That is, we have \[
{(}\int _{N}\mu \phi ^{*}(\xi ),\int _{N}\mu \phi ^{*}(\eta ){)}=\int _{N}\mu \phi ^{*}(\{\xi ,\eta \})\]
 for \( \xi ,\eta \in {\mathcal{C}}^{\cdot }(M) \) and a superfield
\( \phi  \). Moreover, if \( Q \) has hamiltonian \( \Theta  \),
then \( \check{Q} \) has hamiltonian \( \mu _{*}\textrm{ev}^{*}\Theta  \).
This gives us the interaction term of our master action: \[
{\mathbf{S}}_{\textrm{int}}[\phi ]=\int _{N}\mu \phi ^{*}(\Theta )\]
 To see that \( \hat{D} \) also preserves \( {\omega } \), observe
that, if \( \mu  \) is \( D \)-invariant, then \( L_{\hat{D}}\mu _{*}\textrm{ev}^{*}=0 \).
This is essentially because the evaluation map is invariant under
the diagonal action of \( \textrm{Diff}(N) \). Moreover, if \( \omega =d\alpha  \)
for some 1-form \( \alpha  \) on \( M \), this implies that up to
a sign, the hamiltonian of \( \hat{D} \) is \( \iota _{\hat{D}}{\alpha } \),
where \( {\alpha }=\mu _{*}\textrm{ev}^{*}\alpha  \). This gives
the kinetic term \( {\mathbf{S}}_{\textrm{kin}} \) of the master
action. 

To see what this all looks like written out explicitly in coordinates,
we begin with the following general remark. In any odd symplectic
manifold \( {\mathcal{P}} \) with \( \Omega  \) of degree \( -1 \),
the ideal generated by functions of negative degree corresponds to
a \emph{Lagrangian} submanifold \( {\mathcal{L}} \), and in fact
\( {\mathcal{P}} \) is canonically isomorphic to \( T^{*}[-1]{\mathcal{L}} \).
In our case this \( {\mathcal{L}} \) is the space of fields (including
ghosts). It contains the space of classical fields \( {\mathcal{L}}_{0} \),
corresponding to the ideal generated by functions of nonzero degree
(ghosts and antifields). The restriction of any \( {\mathbf{S}} \)
of ghost number zero to \( {\mathcal{L}} \) depends only on the classical
fields, i.e. is a pullback of a functional \( S \) on \( {\mathcal{L}}_{0} \).
This way we recover the classical action. As the critical points of
\( {\mathbf{S}} \) are the fixed points of \( {\mathbf{Q}} \), we
see that the solutions of the classical field equations for \( S \)
are the dg maps from \( (N,D) \) to \( (M,Q) \).

Let us assume from now on that \( N=T[1]N_{0} \) for some closed
oriented \( (n+1) \)-dimensional \( N_{0} \), with \( D=d \), the
de Rham vector field, and \( \mu  \) the canonical measure. In coordinates
\( d=du^{\nu }\frac{\partial }{\partial u^{\nu }} \), and we denote
the induced differential on superfields by \( {\mathbf{d}} \) instead
of \( \hat{D} \). 

As for the target, let \( \omega  \) be written in Darboux coordinates
as \( \omega =\frac{1}{2}dx^{a}\omega _{ab}dx^{b} \). Here \( \omega _{ab} \)
are constants, and so the degrees of \( x^{a} \) and \( x^{b} \)
must add up to \( n \). We shall choose \( \alpha =\frac{1}{2}x^{a}\omega _{ab}dx^{b} \)
as the primitive of \( \omega  \). 

Now, a map \( \phi :N\rightarrow M \) is parametrized in coordinates
by superfields \( \phi ^{*}(x^{a})=\phi ^{a}=\phi ^{a}(u,du) \).
Then \( \Omega  \) is given by\[
\Omega =\int _{T[1]N_{0}}\mu (\frac{1}{2}\delta \phi ^{a}\omega _{ab}\delta \phi ^{b})=\int _{N_{0}}(\frac{1}{2}\delta \phi ^{a}\omega _{ab}\delta \phi ^{b})^{\textrm{top}}\]
 To find the field-antifield splitting and compute the normal form
for \( \Omega  \), let us further keep track of the degree of \( x^{a} \)
by writing it as a subscript: \( x_{i}^{a} \) is of degree \( 0\leq i\leq n \),
then we have \[
\phi _{i}^{a}=\sum _{j=0}^{n+1}\phi _{i,j}^{a}\]
 where \( \phi _{i,j}^{a}=\phi _{i,j}^{a}(u)(du)^{j}=\frac{1}{j!}\phi ^{a}_{i,j,\nu _{1}\ldots \nu _{j}}(u)du^{\nu _{1}}\cdots du^{\nu _{j}} \)
is the \( j \)-form component of \( \phi _{i}^{a} \) whose coefficients
\( \phi _{i,j}^{a}(u) \) have therefore ghost number \( i-j \).
It is easy to see then that we must set, for each field \( \phi _{i,j}^{a} \)
with \( i-j\geq 0 \), its conjugate antifield to be \( \phi ^{\dagger ,i,j}_{a}=(-1)^{ni}\phi _{n-i,n+1-j}^{b}\omega _{ba} \).
Then we can rewrite \( \Omega  \) as \[
\Omega =\int _{N_{0}}(-1)^{i}\sum _{i-j\geq 0}\delta \phi ^{\dagger ,i,j}_{a}\delta \phi ^{a}_{i,j}\]
 The master action is given by \[
{\mathbf{S}}[\phi ]=\int _{T[1]N_{0}}\mu (\frac{1}{2}\phi ^{a}\omega _{ab}{\mathbf{d}}\phi ^{b}+(-1)^{n+1}\phi ^{*}\Theta )\]
 The classical action \( S \) is then recovered by setting all the
antifields to zero. The sign in front of the interaction term is chosen
so that the solutions of the classical field equations \( \delta S=0 \)
coincided with dg maps \( \phi :(T[1]N_{0},d)\rightarrow (M,Q=\{\Theta ,\cdot \}) \).
As we have remarked, \( \mu  \) is invariant under all orientation-preserving
diffeomorphisms of \( N_{0} \), hence \( S \) yields a topological
field theory.

\section{Courant algebroids and the topological closed membrane.}

Specializing the above construction to various choices of the target
one gets many interesting topological field theories. For instance,
in case \( n=1 \) the target is necessarily of the form \( M=T^{*}[1]M_{0} \)
for some manifold \( M_{0} \), and the interaction term is given
by a Poisson bivector field \( \pi  \) on \( M_{0} \). The BV quantization
of the resulting two-dimensional TFT -- the \emph{Poisson sigma model}
-- was extensively studied by Cattaneo and Felder \cite{CF1}\cite{CF2}. 

Here we would like to consider the case \( n=2 \). Symplectic nonnegatively
graded manifolds \( (M,\omega ) \) with \( \textrm{deg}(\omega )=2 \)
were shown in \cite{Roy4-GrSymp} to correspond to vector bundles
\( E\rightarrow M_{0} \) with a fiberwise nondegenerate symmetric
inner product \( <\cdot ,\cdot > \) (of arbitrary signature). The
construction is as follows. Recall that \( \textrm{deg}(M)\leq 2 \),
hence \( M \) fits into a tower of fibrations \[
M=M_{2}\rightarrow M_{1}\rightarrow M_{0}\]
 where \( M_{1} \) is of the form \( E[1] \) for some vector bundle
\( E\rightarrow M_{0} \). Restricting the Poisson bracket to \( M_{1} \)
gives the inner product. Conversely, given \( E \), \( M \) is obtained
as the symplectic submanifold of \( T^{*}[2]E[1] \) corresponding
to the isometric embedding \( E\hookrightarrow E\oplus E^{*} \)with
respect to the canonical inner product on \( E\oplus E^{*} \). If
\( \{x^{i}\} \) are local coordinates on \( M_{0} \) and \( \{e_{a}\} \)
is a local basis of sections of \( E \) such that \( <e_{a},e_{b}>=g_{ab}=\textrm{const}. \),
we get Darboux coordinates \( \{q^{i},p_{i},\xi ^{a}\} \) on \( M \)
(of degrees \( 0 \), \( 2 \) and \( 1 \), respectively), so that
\[
\omega =dp_{i}dq^{i}+\frac{1}{2}d\xi ^{a}g_{ab}d\xi ^{b}=d(p_{i}dq^{i}+\frac{1}{2}\xi ^{a}g_{ab}d\xi ^{b})\]
 Notice that the quadratic hamiltonians \( {\mathcal{C}}^{2}(M) \)
form a Lie algebra under the Poisson bracket, which is isomorphic
to the Lie algebra of infinitesimal bundle automorphisms of \( E \)
preserving \( <\cdot ,\cdot > \). 

It was further shown in \cite{Roy4-GrSymp} that solutions \( \Theta \in {\mathcal{C}}^{3}(M) \)
of the Maurer-Cartan equation \( \{\Theta ,\Theta \}=0 \) correspond
to \emph{Courant algebroid} structures on \( (E,<\cdot ,\cdot >) \).
Such a structure is given by a bilinear operation \( \circ  \) on
sections of \( E \). The condition on \( \circ  \) is that for every
section \( e \) of \( E \), \( e\circ  \) acts by infinitesimal
automorphisms of \( (E,<\cdot ,\cdot >,\circ ) \). In particular,
\( e\circ  \) is a first-order differential operator whose symbol
is a vector field on \( M_{0} \) which we denote by \( a(e) \).
This gives rise to the anchor map \( a:E\rightarrow TM_{0} \). Furthermore,
\( e\circ  \) preserves \( <\cdot ,\cdot > \): \[
a(e)<e_{1},e_{2}>=<e\circ e_{1},e_{2}>+<e_{1},e\circ e_{2}>\]
 as well as the operation \( \circ  \) itself: \[
e\circ (e_{1}\circ e_{2})=(e\circ e_{1})\circ e_{2}+e_{1}\circ (e\circ e_{2})\]
 i.e. \( \circ  \) defines a Leibniz algebra on sections of \( E \).
It follows also that the anchor \( a \) induces a homomorphism of
Leibniz algebras. The only additional property of \( \circ  \) concerns
its symmetric part, namely \[
<e,e_{1}\circ e_{2}+e_{2}\circ e_{1}>=a(e)<e_{1},e_{2}>\]

Now, if we introduce local coordinates as above, the corresponding
\( \Theta \in {\mathcal{C}}^{3}(M) \) will be \[
\Theta =\xi ^{a}P_{a}^{i}(q)p_{i}-\frac{1}{6}T_{abc}(q)\xi ^{a}\xi ^{b}\xi ^{c}\]
 where \( P^{i}_{a} \) is the anchor matrix and \( T_{abc}=<e_{a}\circ e_{b},e_{c}> \).
The Maurer-Cartan equation \( \{\Theta ,\Theta \}=0 \) is equivalent
to the defining properties of \( \circ  \). The corresponding differential
\( Q \) sends a function \( f\in {\mathcal{C}}^{0}(M)=C^{\infty }(M_{0}) \)
to \( a^{*}df \), and a section \( e\in \Gamma (E)={\mathcal{C}}^{1}(M) \)
to \( e\circ \in {\mathcal{C}}^{2}(M) \). 

Now we can write down the sigma-model. Fix a closed oriented 3-manifold
\( N_{0} \), with coordinates \( \{u^{\mu }\} \). The classical
fields are the degree-preserving maps \( T[1]N_{0}\rightarrow M \),
consisting of a smooth map \( X:N_{0}\rightarrow M_{0} \) (the membrane
world-volume), an \( X^{*}E \)-valued 1-form \( A \) and an \( X^{*}T^{*}M_{0} \)-valued
2-form \( F \)%
\footnote{Strictly speaking, this is misleading as the transformation law for
\( p_{i} \) is nonlinear, containing a term quadratic in the \( \xi  \)'s;
as a graded manifold, \( M \) is isomorphic to \( E[1]\oplus T^{*}[2]M_{0} \)
only after a \( g \)-preserving connection has been fixed. This issue
complicates a coordinate-free description of the sigma model.
}. The superfields are written as follows: \[
\begin{array}{rcl}
{\mathbf{q}}^{i} & = & X^{i}+F^{i}_{\dagger }du+\alpha ^{i}_{\dagger }(du)^{2}+\gamma ^{i}_{\dagger }(du)^{3}\\
{\xi }^{a} & = & \beta ^{a}+A^{a}du+g^{ab}A^{\dagger }_{b}(du)^{2}+g^{ab}\beta _{b}^{\dagger }(du)^{3}\\
{\mathbf{p}}_{i} & = & \gamma _{i}+\alpha _{i}du+F_{i}(du)^{2}+X_{i}^{\dagger }(du)^{3}
\end{array}\]
 In particular, the \( X^{*}E \)-valued scalar \( \beta  \) and
the \( X^{*}T^{*}M_{0} \)-valued 1-form \( \alpha  \) are the ghosts,
while the \( X^{*}T^{*}M_{0} \)-valued scalar \( \gamma  \) is the
ghost for ghost. The master action \[
{\mathbf{S}}=\int _{T[1]N_{0}}\mu ({\mathbf{p}}_{i}{\mathbf{d}}{\mathbf{q}}^{i}+\frac{1}{2}{\xi }^{a}g_{ab}{\mathbf{d}}{\xi }^{b}-\Theta ({\mathbf{q}},{\xi },{\mathbf{p}}))\]
 decomposes as \( {\mathbf{S}}=S_{0}+S_{1}+S_{2}+S_{3} \) where the
subscript denotes the number of antifields. Thus, \( S_{0} \) is
the classical action:\[
S_{0}=\int _{N_{0}}F_{i}dX^{i}+\frac{1}{2}A^{a}g_{ab}dA^{b}-A^{a}P_{a}^{i}(X)F_{i}+\frac{1}{6}T_{abc}(X)A^{a}A^{b}A^{c}\]
 and the rest of the terms are as follows:\[
\begin{array}{rcl}
S_{1} & = & \int _{N_{0}}(-\beta ^{a}P_{a}^{i}X_{i}^{\dagger }+(d\beta ^{c}-g^{ac}P_{a}^{i}\alpha _{i}+\beta ^{a}A^{b}T_{abr}g^{rc})A^{\dagger }_{c}+\\
 &  & +(-d\alpha _{j}-\beta ^{a}\partial _{j}P^{i}_{a}F_{i}-A^{a}\partial _{j}P^{i}_{a}\alpha _{i}+\frac{1}{2}\partial _{j}T_{abc}\beta ^{a}A^{b}A^{c})F^{j}_{\dagger }+\\
 &  & +(-d\gamma _{j}-\beta ^{a}\partial _{j}P^{i}_{a}\alpha _{i}-A^{a}\partial _{j}P^{i}_{a}\gamma _{i}+\frac{1}{2}\partial _{j}T_{abc}\beta ^{a}\beta ^{b}A^{c})\alpha ^{j}_{\dagger }+\\
 &  & +(\frac{1}{2}T_{abr}g^{rc}\beta ^{a}\beta ^{b}-g^{ac}P_{a}^{i}\gamma _{i})\beta ^{\dagger }_{c}+(\frac{1}{6}\partial _{j}T_{abc}\beta ^{a}\beta ^{b}\beta ^{c}-\beta ^{a}\partial _{j}P^{i}_{a}\gamma _{i})\gamma ^{j}_{\dagger })
\end{array}\]
 \[
\begin{array}{rcl}
S_{2} & = & \int _{N_{0}}((\frac{1}{2}\partial _{j}T_{abr}g^{rc}\beta ^{a}\beta ^{b}-g^{ac}\partial _{j}P^{i}_{a}\gamma _{i})F_{\dagger }^{j}A^{\dagger }_{c}-\frac{1}{2}(\beta ^{a}\partial _{j}\partial _{k}P^{i}_{a}\alpha _{i}+A^{a}\partial _{j}\partial _{k}P^{i}_{a}\gamma _{i}-\\
 &  & -\frac{1}{2}\partial _{j}\partial _{k}T_{abc}\beta ^{a}\beta ^{b}A^{c})F^{j}_{\dagger }F^{k}_{\dagger }-(\beta ^{a}\partial _{j}\partial _{k}P^{i}_{a}\gamma _{i}-\frac{1}{6}\partial _{j}\partial _{k}T_{abc}\beta ^{a}\beta ^{b}\beta ^{c})F^{j}_{\dagger }\alpha _{\dagger }^{k})
\end{array}\]
 \[
S_{3}=\int _{N_{0}}\frac{1}{6}(-\beta ^{a}\partial _{j}\partial _{k}\partial _{l}P^{i}_{a}\gamma _{i}+\frac{1}{6}\partial _{j}\partial _{k}\partial _{l}T_{abc}\beta ^{a}\beta ^{b}\beta ^{c})F^{j}_{\dagger }F^{k}_{\dagger }F^{l}_{\dagger }\]
 It appears our sigma model has very complicated 2-algebroid gauge
symmetries generated by parameters \( \alpha _{i},\beta ^{a} \) and
\( \gamma _{i} \); writing down the master action without the help
of AKSZ would have been extremely difficult. It would be desirable
to better understand the structure of the gauge symmetries.

In conclusion, let us point out some special cases.

\begin{example}
Let \( M_{0}=\{\textrm{pt}\} \). Then \( (E,<\cdot ,\cdot >) \)
is just a vector space with an inner product. A Courant algebroid
structure reduces to that of a quadratic Lie algebra with structure
constants \( T_{abc}=<[e_{a},e_{b}],e_{c}> \). A quick glance reveals
that in this case \( S_{0} \) is the classical Chern-Simons functional,
for which the master action was written down in \cite{AKSZ}. 
\end{example}

\begin{example}
Let \( M=T^{*}[2]T^{*}[1]M_{0}=T^{*}[2]T[1]M_{0} \), with coordinates
\( \{q^{i},\xi ^{i},\theta _{i},p_{i}\} \) of degree 0,1,1 and 2,
respectively (one thinks \( \xi ^{i}=dx^{i} \)). Then \( \omega =dp_{i}dq^{i}+d\xi ^{i}d\theta _{i}=d(p_{i}dq^{i}+\xi ^{i}d\theta _{i}) \),
and we consider \( \Theta =\xi ^{i}p_{i}-\frac{1}{6}c_{ijk}(q)\xi ^{i}\xi ^{j}\xi ^{k} \),
where \( c=\frac{1}{6}c_{ijk}(q)\xi ^{i}\xi ^{j}\xi ^{k} \) is a
3-form on \( M_{0} \). Clearly \( \Theta  \) obeys Maurer-Cartan
if and only if \( dc=0 \). The corresponding Courant algebroid structure
on \( E=TM\oplus T^{*}M \) (with the canonical inner product) is
given by \[
(X+\xi )\circ (Y+\eta )=[X,Y]+L_{X}\eta -\iota _{Y}d\xi +\iota _{X\wedge Y}c\]
 The classical fields for the corresponding topological membrane action
are comprised of the membrane world-volume \( X:N_{0}\rightarrow M_{0} \),
an \( X^{*}TM_{0} \)-valued 1-form \( A \), an \( X^{*}T^{*}M_{0} \)-valued
1-form \( B \) and an \( X^{*}T^{*}M_{0} \)-valued 2-form \( F \).
The classical action \[
S_{0}[X,A,B,F]=\int _{N_{0}}F_{i}dX^{i}+A^{i}dB_{i}-A^{i}F_{i}+\frac{1}{6}c_{ijk}(X)A^{i}A^{j}A^{k}\]
 was considered by Park \cite{Park}. We leave it to the reader to
write down the master action in this case.
\end{example}


\begin{thebibliography}{10}

\bibitem{AKSZ}
M.~Alexandrov, M.~Kontsevich, A.~Schwarz, and O.~Zaboronsky.
\newblock The geometry of the {Master} equation and topological quantum 
field
  theory.
\newblock {\em Int. J. Modern Phys. A}, 12(7):1405--1429, 1997.

\bibitem{BV1}
I.~Batalin and G~Vilkovisky.
\newblock Gauge algebra and quantization.
\newblock {\em Phys. Lett.}, 102B:27, 1981.

\bibitem{CF1}
A.~Cattaneo and G.~Felder.
\newblock A path integral approach to the {Kontsevich} quantization 
formula.
\newblock {\em Commun. Math. Phys.}, 212:591--611, 2000.

\bibitem{CF2}
A.~Cattaneo and G.~Felder.
\newblock On the {AKSZ} formulation of the {Poisson} sigma model.
\newblock {\em Lett. Math. Phys.}, 56:163--179, 2001.

\bibitem{HP2}
C.~Hofman and J.-S. Park.
\newblock {BV} quantization of topological open membranes.
\newblock preprint hep-th/0209214, 2002.

\bibitem{HP1}
C.~Hofman and J.-S. Park.
\newblock Topological open membranes.
\newblock preprint hep-th/0209148, 2002.

\bibitem{Ikeda}
N.~Ikeda.
\newblock {Chern-Simons} gauge theory coupled with {BF} theory.
\newblock {\em Int. J. Mod. Phys.}, A18:2689--2702, 2003.
\newblock hep-th/0203043.

\bibitem{Park}
J.-S. Park.
\newblock Topological open $p$-branes.
\newblock preprint hep-th/0012141, 2000.

\bibitem{Roy4-GrSymp}
D.~Roytenberg.
\newblock On the structure of graded symplectic supermanifolds and 
{Courant}
  algebroids.
\newblock In Theodore Voronov, editor, {\em Quantization, Poisson Brackets 
and
  Beyond}, volume 315 of {\em Contemp. Math.} Amer. Math. Soc., 
Providence, RI,
  2002.
\newblock math.SG/0203110.

\bibitem{Roy3-QuasiLie}
D.~Roytenberg.
\newblock {Quasi-Lie} bialgebroids and twisted {Poisson} manifolds.
\newblock {\em Lett. Math. Phys.}, 61(2):123--137, 2002.
\newblock math.QA/0112152.

\bibitem{SchStr}
P.~Schaller and T.~Strobl.
\newblock Poisson structure induced (topological) field theories.
\newblock {\em Mod. Phys. Lett.}, A9:3129--3136, 1994.
\newblock hep-th/9405110.

\bibitem{Vor3}
T.~Voronov.
\newblock Graded manifolds and {Drinfeld} doubles for {Lie} bialgebroids.
\newblock In Theodore Voronov, editor, {\em Quantization, Poisson Brackets 
and
  Beyond}, volume 315 of {\em Contemp. Math.} Amer. Math. Soc., 
Providence, RI,
  2002.
\newblock Preprint math.DG/0105237.

\end{thebibliography}

\end{document}